\documentclass[preprint2]{aastex}

\slugcomment{Received 3 Jan 2016, Accepted on 2 Mar 2016}

\shorttitle{{\it Suzaku} observation of SN2014J}
\shortauthors{Terada et al.}

\begin{document}


\title{Measurements of the Soft Gamma-ray Emission from SN2014J with {\it Suzaku}}


\author{Y. Terada\altaffilmark{1}, 
K. Maeda\altaffilmark{2,3}, 
Y. Fukazawa\altaffilmark{4},
A. Bamba\altaffilmark{5}, 
Y. Ueda\altaffilmark{2},
S. Katsuda\altaffilmark{6},
T. Enoto\altaffilmark{2},
T. Takahashi\altaffilmark{6},
T. Tamagawa\altaffilmark{7}, 
F. K. R{\"o}pke\altaffilmark{8,9},
A. Summa\altaffilmark{10},   
R. Diehl\altaffilmark{11}
}

\altaffiltext{1}{Graduate School of Science and Engineering, 
Saitama University, 255 Shimo-Ohkubo, Sakura, Saitama 338-8570,Japan}
\altaffiltext{2}{Department of Astronomy, Kyoto University, 
Kitashirakawa-Oiwake-cho, Sakyo-ku, Kyoto 606-8502, Japan}
\altaffiltext{3}{Kavli Institute for the Physics and Mathematics of the Universe (WPI), 
University of Tokyo, 5-1-5 Kashiwanoha, Kashiwa, Chiba 277-8583, Japan}
\altaffiltext{4}{Department of Physical Science, Hiroshima University, 
1-3-1 Kagamiyama, Higashi-Hiroshima, Hiroshima 739-8526, Japan}
\altaffiltext{5}{Department of Physics and Mathematics,
Aoyama Gakuin University,
5-10-1 Fuchinobe Chuo-ku, Sagamihara, Kanagawa 252-5258, Japan}
\altaffiltext{6}{Institute of Space and Astronautical Science,
Japan Aerospace eXploration Agency,
3--1--1 Yoshinodai, Sagamihara, Kanagawa 229--8510, Japan}
\altaffiltext{7}{RIKEN, 2-1 Hirosawa, Wako, Saitama 351-0198, Japan}
\altaffiltext{8}{Zentrum f{\"u}r Astronomie der Universit{\"a}t Heidelberg, Institut f{\"u}r Theoretische Astrophysik, Philosophenweg 12, 69120 Heidelberg, Germany}
\altaffiltext{9}{Heidelberger Institut f{\"u}r Theoretische Studien, Schloss-Wolfsbrunnenweg 35, 69118 Heidelberg, Germany}
\altaffiltext{10}{Max-Planck-Institut f{\"u}r Astrophysik, Karl-Schwarzschild-Str. 1, D-85748 Garching, Germany}
\altaffiltext{11}{Max-Planck-Institut f{\"u}r extraterrestrische Physik, 85741, Garching, Germany}


\begin{abstract}
The hard X-ray detector (HXD) onboard {\it Suzaku} measured
soft $\gamma$-rays from the Type Ia supernova SN2014J 
at $77\pm2$ days after the explosion.
Although the confidence level of the signal is 
about 90\% (i.e., $2 \sigma$),
the $3 \sigma$ upper limit has been derived at
$< 2.2 \times10^{-4}$ ph s$^{-1}$ cm$^{-2}$ in the 170 -- 250 keV band
as the first independent measurement of soft $\gamma$-rays 
with an instrument other than {\it INTEGRAL}.
For this analysis, we have examined the reproducibility of the NXB model 
of HXD/GSO using blank sky data. 
We find that the residual count rate in the 90 -- 500 keV band is 
distributed around an average of 0.19\% with a standard deviation of 
0.42\% relative to the NXB rate. 
The averaged residual signals are consistent with 
that expected from the cosmic X-ray background.
The flux of SN2014J derived from {\it Suzaku} measurements 
taken in one snapshot at $t=77\pm2$ days 
after the explosion is consistent with the {\it INTEGRAL} values 
averaged over the period between $t=$50 and 100 days and
also with explosion models of single or double degenerate scenarios.
Being sensitive to the total ejecta mass surrounding the radioactive material,
the ratio between continuum and line flux in the soft gamma-ray regime 
might distinguish different progenitor models.
The {\it Suzaku} data have been examined with this relation at $t=77\pm2$ days,
but could not distinguish models between single and 
double degenerate-progenitors.
We disfavor explosion models with larger $^{56}$Ni masses than 1 $M_\odot$,
from our $1 \sigma$ error on the 170-250 keV X-ray flux of
$(1.2\pm0.7) \times10^{-4}$ ph s$^{-1}$ cm$^{-2}$.
\end{abstract}





\section{Introduction}
\label{section:intro}
Type Ia supernovae (SNe) are very bright stellar explosions 
which are detectable at optical wavelengths across cosmological distances. 
It is widely accepted that they originate from thermonuclear explosions 
of carbon-oxygen white dwarfs (WDs) in binary systems. 
They are among the most matured standardizable candles 
\citep{Phillips93,Riess98,Perlmutter99}, 
having a tight but phenomenologically calibrated relation 
between the optical peak luminosity and 
the decline rate of the light curve in the $B$-band.

\begin{deluxetable}{cccccc}
\tabletypesize{\scriptsize}
\tablecaption{{\it Suzaku} Observations towards M82}
\tablewidth{0pt}
\tablehead{
 \colhead{OBSID}  &
 \colhead{Target Name}   &
 \colhead{Date}   &
 \colhead{HXD Exposure} &
 \colhead{PIN Count rate\tablenotemark{a}} &
 \colhead{GSO Count rate\tablenotemark{b}} \\
  & &(yyyy/mm/dd)&(ks)&($10^{-2}$ c/s)&($10^{-1}$ c/s)
}
\startdata
100033010 & M82-Wind & 2005/10/04 &  28.6 & $1.6\pm0.4$ & $1.2\pm1.2$ \\
100033020 & M82-Wind & 2005/10/19 &  36.1 & $2.6\pm0.4$ & $0.0\pm1.1$ \\
100033030 & M82-Wind & 2005/10/28 &  24.0 & $3.2\pm0.5$ & $0.2\pm0.4$ \\
702026010 & M82 X-1  & 2007/09/24 &  28.4 & $2.6\pm0.4$ & $0.0\pm0.4$ \\
908005010\tablenotemark{c} & SN 2014J & 2014/03/30 & 193.9 & $2.6\pm0.1$ & $1.8\pm0.1$ \\
\enddata
\tablenotetext{a}{Count rate of NXB-and-CXB subtracted signals 
of the HXD PIN in the 13 -- 70 keV band.}
\tablenotetext{b}{Count rate of NXB subtracted signals 
of the HXD GSO in the 90 -- 500 keV band.}
\tablenotetext{c}{{\it Suzaku} observation in 2014 
defined as 'OBS2014' in the text.}
\label{table:observation_list}
\end{deluxetable}

However, the progenitors of Type Ia SNe have been 
poorly constrained observationally 
despite many on-going attempts \cite[see e.g.,][for reviews]{Maoz14}. 
There are several variants in terms of the ignition and propagation 
of the thermonuclear flame \citep{Hillebrandt00}, 
which can have different characteristics in 
(1) the evolution toward the explosion, and 
(2) in the mass of the exploding WD. 
The evolution scenarios are roughly divided into two categories 
referring to the nature of the progenitor systems;
the single degenerate scenario 
\citep[hereafter SD;][]{Whelan73,Nomoto82} 
(a C+O WD and a main-sequence/red-giant companion) or 
double degenerate scenario 
\citep[hereafter DD;][]{Iben84,Webbink84} (a merger of two C+O WDs). 
The mass of the exploding WD(s) is linked to the progenitor systems
and their evolution scenario,
which would affects on the cosmological usage of SN Ia as distance indicators.
In the SD scenario the most popular model involves a Chandrasekhar-mass WD 
\citep[e.g.,][]{Nomoto82}. 
The original DD scenario is also associated with 
the Chandrasekhar-mass WD \citep[e.g.,][]{Iben84}. 
In a recently proposed variant of the DD model, 
the so-called violent merger model \citep{Pakmor10,Roepke12}, 
the total mass of the ejecta (i.e., a sum of the two WDs) can exceed 
the Chandrasekhar-mass limit, 
a specific model of which is for example presented in \citet{Summa13}. 
Determining the ejecta mass and/or the progenitor WD is 
therefore of particular importance \citep[e.g.,][]{Scalzo14,Yamaguchi15,Katsuda15}.

As demonstrated in the optical light curves of SNe Ia, 
they produce a large amount of $^{56}$Ni in the explosion, 
on average $\sim 0.6 M_{\odot}$. 
Direct measurements of $\gamma$-ray emission from the decay chain, 
$^{56}$Ni $\rightarrow ^{56}$Co $\rightarrow ^{56}$Fe \citep{Arnett79}, 
have been suggested to provide not only the direct evidence 
for the thermonuclear nature of SNe Ia \citep{Ambwani88,Milne04} 
but also various diagnostics to discriminate different models 
\citep[e.g., see][for predictions based on multi-dimensional explosion models]{Maeda12,Summa13}. 
Among various possibilities, 
it has been suggested to be a strong probe to 
the mass of the explosion systems \citep{Sim08,Summa13}, 
i.e., either a single Chandrasekhar-mass WD or merging two WDs 
for which the total mass can exceed the Chandrasekhar-mass.

Despite the strong motivation to analyze 
the $\gamma$-ray emission from SNe Ia, 
no solid detection had been reported until 2014, 
including attempts for SN1991T \citep{Lichti94, Leising95}, 
SN1998bu \citep{Georgii02} and SN2011fe \citep{Isern13}. 
The situation changed in 2014, after 
SN2014J was discovered on 22 January 2014 \citep{Fossey14} 
in the nearby star-burst galaxy M82 
at the distance $d \sim 3.5$ Mpc \citep{Dalcanton09, Karachentsev06} 
and was classified as the closest Type Ia SN \citep{Ayani14,Cao14,Itoh14} 
in the last three decades. 
The reconstructed date of the explosion was 14.75 January 2014 \citep{Zheng14}. 
In the MeV $\gamma$-ray band, the {\it INTEGRAL} satellite made possible
the first detection of $^{56}$Co $\rightarrow ^{56}$Fe lines 
at 847 and 1238 keV 
at $(2.34 \pm 0.74) \times 10^{-4}$ ph cm$^{-2}$ s$^{-1}$ and 
$(2.78 \pm 0.74) \times 10^{-5}$ ph cm$^{-2}$ s$^{-1}$, respectively, 
in an average of the 50 to 100 days after the explosion 
\citep{Churazov14,Diehl15}. 
Even at earlier phases of 20 days after the explosion, 
the detection of $^{56}$ Ni $\rightarrow ^{56}$Co lines 
at 152 and 812 keV at $(1.10 \pm 0.42) \times 10^{-4}$ ph cm$^{-2}$ s$^{-1}$ and 
$(1.90 \pm 0.66) \times 10^{-5}$ ph cm$^{-2}$ s$^{-1}$, 
respectively, was reported \citep{Diehl14}.
Analysing the time evolution of $^{56}$Co lines \citep{Diehl15, Siegert15}, 
a $^{56}$Ni mass of 0.5 $M_\odot$ was derived.
But a clear discrimination of models between SD and DD
does not seem to be possible, 
both from limitations of the measured $\gamma$-ray intensity evolution 
and the theoretical prediction from different models.

These studies provided the first detection of 
nuclear $\gamma$-ray emission from SNe Ia, 
and indeed the only detection of nuclear $\gamma$-ray emission 
from objects beyond the local group of galaxies. 
This detection relies on the SPI and IBIS instruments 
on the same satellite {\it INTEGRAL}, and 
additional confirmation by a fully independent instrument is important. 
Moreover, 
while these previous reports mostly focused on the detection of the lines, 
a wealth of additional information is contained 
in the continuum emission. 
The MeV decay lines are scattered down to lower energy by Compton scattering,
creating continuum emission above $\sim 100$ keV 
\citep[e.g.,][]{Ambwani88,Sim08,Summa13}.
This process is more important for more dense ejecta,
unlike the line strengths which become weaker for more dense ejecta.
Therefore, combining the information from the lines and the continuum, 
one expects to obtain additional insight into the properties of 
the SN ejecta that is then linked to the progenitor star. 
Indeed, the detection of continuum in the energy range of $200 - 400$ keV 
by INTEGRAL was reported by \citet{Churazov14}. 
In this paper, we report a measurement of
the $\gamma$-ray continuum from SN 2014J 
with the {\it Suzaku} X-ray satellite \citep{Mitsuda07}.
In sections \ref{section:observation} and \ref{section:analysis} and 
we test several explosion models to constrain 
the mass of $^{56}$Ni and the mass of the exploding WD system
in section \ref{section:discussion}.

\section{Observation and Data Reduction}
\label{section:observation}
\subsection{ToO Observation with {\it Suzaku}}
\label{section:observation:obs}
The X-ray satellite {\it Suzaku} carries two active X-ray instruments 
onboard \citep{Mitsuda07}; 
the X-ray Imaging Spectrometer \citep[XIS; ][]{Koyama07} and 
the hard X-ray detector \citep[HXD;][]{Takahashi07} 
to observe the 0.2 -- 12 keV and the 13 -- 600 keV bands, respectively. 
The HXD is a hybrid detector with PIN-type Si photo-diodes 
for the 13 -- 70 keV band 
and phoswitch-type scintillation counters using 
Gd$_2$SiO$_5$ (hereafter GSO) crystals surrounded by 
Bi$_4$Ge$_3$O$_{12}$ (hereafter BGO) crystals 
for the 60 -- 600 keV band \citep{Takahashi07}. 
It has a comparable or better sensitivity than 
that of {\it INTEGRAL} instruments in the 60 -- 200 keV band 
on an 'one-shot' short observation and therefore, 
it is suitable for our purpose to independently detect 
the soft $\gamma$-ray emission from SN2014J.

We triggered ToO observation of SN2014J with {\it Suzaku} from 
2014 March 30 12:18 UT to 3 April 17:23 UT (OBSID=908005010), 
which is about $t = 77 \pm 2$ days after the explosion of SN2014J, 
soon after the day when the sun angle allows the satellite operation. 
The target position was set to ($\alpha, \delta$)[J2000] = 
($09^{\rm h}55^{\rm m}42.12^{\rm s}$, $+69^{\circ}40^{\prime}26.0^{"}$) 
at the XIS nominal pointing position. 
The HXD was operated in the nominal mode; 
the bias voltages for one half of 64 PIN diodes were operated at 400 V and 
the other half at 500 V, and the photo-multipliers for scintillators were 
operated in the nominal setting of the high voltages.
We also used previous observations towards the M82 region before the
explosion of SN2014J for comparison in later sections.
The observation in 2014J with OBSID=908005010 (hereafter OBS2014)
and previous ones are summarized in Table \ref{table:observation_list}.

\begin{deluxetable}{lcc}
\tabletypesize{\scriptsize}
\tablecaption{Systematic error of HXD-GSO NXB model for OBS2014}
\tablewidth{0pt}
\tablehead{
 \colhead{ID}&
 \colhead{Energy band (keV)}  &
 \colhead{Reproducibility (\%) \tablenotemark{a}}
}
\startdata
{\it i}   & 86 - 120 & 0.48 \\
{\it ii}  &120 - 144 & 0.55 \\
{\it iii} &144 - 176 & 2.37 \\
{\it iv}  &176 - 202 & 0.01 \\
{\it v}   &202 - 256 & 0.05 \\
{\it vi}  &256 - 342 & 0.55 \\
{\it vii} &342 - 500 & 1.80 \\
\enddata
\tablenotetext{a}{Reproducibility of the non X-ray background model
defined by the percentage between count rates of the residual and 
the NXB model.}
\label{table:nxb_systematic_sn2014j}
\end{deluxetable}

\subsection{Data Reduction}
\label{section:observation:obs}

The observation data-sets were processed 
by the standard {\it Suzaku} pipeline 
version 2.8.20.35, with the calibration version (CALDBVER) of 
hxd20110913, xis20121106, xrt20110630, and xrs20060410 for OBS2014.
In the analysis of other OBSIDs in the following Section \ref{section:analysis},
all the data are reprocessed by the \emph{ftool}, 'aepipeline', 
with the latest CALDB files with equivalent version of OBS2014.
Spectral fitting was performed with XSPEC version 12.8.1g 
in HEADAS 6.15.1 package.
Background was estimated from models for instrumental
(i.e., 'non-X-ray') background plus cosmic diffuse X-ray background,
both fitted to the SN2014J and other independent data (see below).

We did not use the XIS data, because bright X-rays 
from the ultra-luminous source M82 X-1 strongly contaminated 
the SN2014J region.

Cleaned event lists of the HXD are obtained by 
the standard selection criteria.
The net exposure for the HXD is 193.9 ks. 
The non-X-ray background (NXB) is estimated 
using the methods described in \citet{Fukazawa09}.
We used the NXB events of both PIN and GSO with 
METHOD=''LCFITDT (bgd\_d)'' and the version of METHODV=2.6ver1110-64.
Here, if we subtract NXB events from OBS2014 data,
the net count rates of PIN and GSO are
$(2.6\pm0.1) \times 10^{-1}$ c s$^{-1}$ and 
$(1.8\pm0.1) \times 10^{-2}$ c s$^{-1}$, respectively, 
in the 13 -- 70 or the 90 -- 500 keV bands, respectively.
Count rates for other observations towards M82 than OBS2014 are 
also summarized in Table \ref{table:observation_list}.

On the HXD PIN detector, the count rate of OBS2014 
in Table \ref{table:observation_list} shows no significant excess 
over the others. According to \citet{Miyawaki09}, 
most of PIN signals can be considered as hard X-rays from 
the ULX M82 X-1, whereas $\gamma$-rays from Type Ia SNe should be weak 
in this energy band below 100 keV \citep{Maeda12,The14}. 
Therefore, in the following sections, we concentrate on checking
the detectability of $\gamma$-rays from SN2014J with the HXD GSO 
in the energy band above 90 keV.

\begin{figure}[h]
\centerline{\includegraphics[angle=0,width=0.45\textwidth]{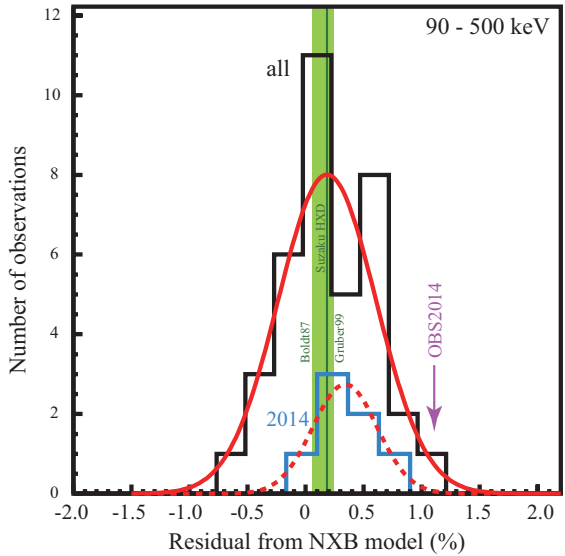}}
\caption{Distributions of residuals of GSO signals of 
blank sky observations (Table \ref{table:list_blanksky}) 
from the NXB models in the 90 -- 500 keV band 
are shown in the histograms.
Distribution for all the 37 observations is shown in black
and that for 7 selected observations taken within 2 months 
before or after the SN2014J observation 
(i.e., OBSIDs in Table\ref{table:list_blanksky} with note $e$) are in cyan.
The best-fit Gaussian models for them are plotted in red.
For comparison, the average value for all the blank sky observations 
(whose spectrum is shown in Figure \ref{fig:spec_sn2014j_cxb} blue) 
is shown in dark green line, and 
the CXB levels by HEAO-1 \citep{Boldt87,Gruber99} 
are shown in the green hatched box.
The residual from the NXB model for OBS2014 is shown in the purple with arrow.}
\label{fig:nxb_systematic_sn2014j}
\end{figure}

\section{Analysis and Results}
\label{section:analysis}
\subsection{Signal Level compared with the Systematics of Non X-ray background}
\label{section:analysis:nxb}
The systematic error is mainly determined by the reproducibility 
of the NXB model, which is about 1 \% or less for GSO in 
more than 10 ks exposure \citep{Fukazawa09}.
For OBS2014, it was confirmed with the earth occultation data 
during the observation,
whose exposure is only 17.7 ks with standard criteria of Cut-off-Rigidity or 
21.5 ks when we do not exclude data with bad conditions of Cut-off-Rigidity
(see Table \ref{table:nxb_systematic_sn2014j}).
We estimated a systematic uncertainties of 0.1 -- 0.6 \% 
(except for one energy bins {\it (iii)} and {\it (vii)} in the table).
Note that the definition of energy bins in 
Table \ref{table:nxb_systematic_sn2014j} is determined by bins 
in the NXB estimation by \citet{Fukazawa09}.
In order to perform a more precise check on 
the reproducibility of the GSO NXB models 
for longer exposures than 21.5 ks 
during the sky observations (i.e., not earth occultation data),
we estimated them with the ``blank sky'' observations for the HXD GSO.
Among all the {\it Suzaku} observations after the launch in 2005 to 2014, 
we first picked up 140 observations whose exposures of the XIS 
exceed 120 ks, and then selected ``blank sky'' observations for GSO 
with the following criteria:
(1) PIN counts in the 50 -- 60 keV or the 60 -- 70 keV bands 
do not exceed 3.5\% of the NXB models, 
(2) GSO counts in the 90 -- 500 keV do not exceed 2.0\% 
of the NXB models, or
the number of energy bands in which GSO counts exceed 1.0\% of each NXB level
is less than half 
(i.e., 3 among 7 bands defined in Table \ref{table:nxb_systematic_sn2014j}),
(3) the systematic errors of GSO NXB models estimated by the earth data 
do not exceed 2 \%.
Finally, we got 37 ``blank sky'' observations 
as listed in Table \ref{table:list_blanksky}. 
The total exposure of them is 4.49 Ms. 
The reproducibility of the GSO NXB model for each observation 
is also listed in the table.
As a result, the reproducibility of NXB models distributes
in $0.19 \pm 0.42$ \% for all the 37 observations
or $0.34 \pm 0.28$ \% for observations near the SN2014J date,
with $1 \sigma$ errors, 
as demonstrated in Figure \ref{fig:nxb_systematic_sn2014j}.
Note that this discrepancy (i.e., 0.19\% offset here) 
between blank sky observation and NXB models comes from 
a contamination of CXB emission in the field of view of the HXD GSO and 
from the Earth's albedo emission included in NXB models;
the effect is seen in green lines of Figure \ref{fig:nxb_systematic_sn2014j} 
and is numerically estimated in Section \ref{section:analysis:background}.
The standard deviation (0.42\%) corresponds to 42\% of 
the NXB-subtracted GSO signals of OBS2014 in the same energy range.

\begin{deluxetable}{clclcc}
\tabletypesize{\scriptsize}
\tablecaption{List of reference observations}
\tablewidth{0pt}
\tablehead{
    \colhead{OBSID}     & \colhead{Target Name} & 
    \colhead{Position\tablenotemark{a}} & 
    \colhead{Obs.\ Date \tablenotemark{b}} &
    \colhead{Exp.\tablenotemark{c}} & \colhead{Res.\tablenotemark{d}} \\
    & & (RA,Dec) & (yyyy/mm/dd) & (ks) & (\%)
  }
\startdata
101012010 & PERSEUS CLUSTER & (49.9436, 41.5175) & 2006/08/29  & 133.2 & -0.044 \\
402015010 & LS 5039 & (276.5633, -14.9109) & 2007/09/09  & 167.7 & 0.429 \\
402033010 & SIGMA GEM & (115.843, 28.9438) & 2007/10/21  & 116.2 & 0.000 \\
404001010 & AE AQUARII & (310.0451, -0.9346) & 2009/10/16  & 126.9 & -0.036 \\
408019020 & V1223 SGR & (283.7576, -31.1629) & 2014/04/10\tablenotemark{e} & 137.3 & 0.464 \\
408024030 & V2301 OPH & (270.1437, 8.1764) & 2014/04/05\tablenotemark{e} & 53.2 & 0.103 \\
408029010 & V1159 ORI & (82.2495, -3.563) & 2014/03/16\tablenotemark{e} & 177.9 & 0.157 \\ 
500010010 & RXJ 0852-4622 NW & (132.2926, -45.6157) & 2005/12/19  & 214.8 & -0.299 \\
502046010 & SN1006 & (225.7268, -41.9424) & 2008/02/25  & 171.4 & 0.347 \\
502048010 & 47 TUCANAE & (6.2112, -71.9961) & 2007/06/10  & 104.8 & 0.161 \\
502049010 & HESS J1702-420 & (255.6874, -42.0709) & 2008/03/25  & 131.4 & -0.076 \\
503085010 & TYCHO SNR & (6.3139, 64.1469) & 2008/08/04  & 269.6 & 0.923 \\
503094010 & SNR 0049-73.6 & (12.7817, -73.3677) & 2008/06/12  & 100.7 & 0.077 \\
506052010 & G352.7-0.1 & (261.9227, -35.1119) & 2012/03/02  & 159.7 & -0.680 \\
507015030 & IC 443 & (94.3026, 22.7461) & 2013/03/31  & 106.3 & 0.700 \\
508003020 & W44 SOUTH & (284.0546, 1.2208) & 2014/04/09\tablenotemark{e} & 27.7 & -0.103 \\
508006010 & W28 SOUTH & (270.2522, -23.558) & 2014/03/22\tablenotemark{e} & 33.5 & 0.690 \\
508017010 & RX J1713.7-3946 NE & (258.6449, -39.4419) & 2014/02/26\tablenotemark{e} & 97.8 & 0.601 \\
508072010 & 0509-67.5 & (77.4163, -67.5163) & 2013/04/11  & 154.2 & 1.006 \\
701003010 & IRAS13224-3809 & (201.327, -38.416) & 2007/01/26  & 158.5 & -0.212 \\
701031010 & MARKARIAN 335 & (1.5539, 20.2624) & 2006/06/21  & 131.7 & 0.138 \\
701047010 & MRK 1 & (19.06, 33.0289) & 2007/01/11  & 117.8 & 0.041 \\
701056010 & PDS 456 & (262.0807, -14.2604) & 2007/02/24  & 164.3 & -0.413 \\
702059010 & 3C 33 & (17.2445, 13.2796) & 2007/12/26  & 99.2 & 0.690 \\
703048010 & PKS 0528+134 & (82.7307, 13.5905) & 2008/09/27  & 126.4 & 0.607 \\
703049010 & 3C279 & (194.0685, -5.7338) & 2009/01/19  & 77.5 & 0.657 \\
704009010 & NGC 454 & (18.511, -55.3853) & 2009/04/29  & 106.0 & 0.491 \\
704062010 & NGC3516 & (166.8656, 72.6213) & 2009/10/28  & 178.2 & 0.578 \\
707035020 & PDS 456 & (262.0805, -14.2617) & 2013/03/03  & 138.1 & 0.090 \\
708016010 & MKN 335 & (1.5767, 20.2085) & 2013/06/11  & 116.6 & 0.140 \\
800011010 & A3376 WEST RELIC & (90.0415, -39.9946) & 2005/11/07  & 105.1 & -0.104 \\
801064010 & NGC 4472 & (187.4441, 8.005) & 2006/12/03  & 96.4 & 0.253 \\
802060010 & ABELL 2029 & (227.4644, 6.0238) & 2008/01/08  & 139.2 & 0.349 \\
803053010 & ABELL S753 RELIC & (211.0241, -34.0331) & 2009/01/07  & 92.3 & 0.874 \\
808043010 & FORNAX A EAST LOBE & (51.0149, -37.2799) & 2013/08/02  & 125.7 & 0.094 \\
808063010 & ESO318-021 & (163.2697, -40.3328) & 2013/12/13  & 125.2 & -0.496 \\
809119010 & ABELL2345EAST & (321.8675, -12.1557) & 2014/04/30\tablenotemark{e} & 83.0 & 0.161 \\
\enddata
\tablenotetext{a}{Target position, R.A. and Dec, in J2000 coordinate. }
\tablenotetext{b}{Observation start date in year/month/day. }
\tablenotetext{c}{Exposure for the HXD in ks.}
\tablenotetext{d}{Residuals of signals from NXB models 
in the 90 -- 500 keV band, shown in the percentage of the NXB.}
\tablenotetext{e}{Guest observations before or after OBS2014.}
\label{table:list_blanksky}
\end{deluxetable}

\begin{figure}[h]
\centerline{\includegraphics[angle=-90,width=0.45\textwidth]{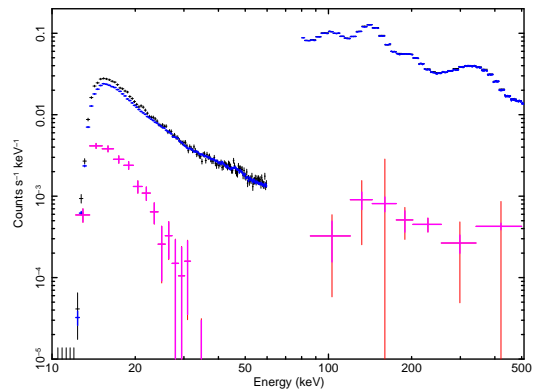}}
\caption{The X-ray spectra in OBS2014 with {\it Suzaku} HXD PIN (below 60 keV)
and GSO (above 80 keV). Crosses in black and blue represent the raw data 
and the NXB models, respectively.
The background subtracted spectra with statistical errors ($1 \sigma$)
are shown in magenta. Similarly, those with systematic errors of 
NXB models are shown in red; the systematic error 
for PIN is set to be 3\% of non X-ray background level \citep{Fukazawa09}
whereas those for GSO data are determined by each channel 
as summarized in Table \ref{table:nxb_systematic_sn2014j}.}
\label{fig:raw_spec_sn2014j}
\end{figure}

\begin{figure}[h]
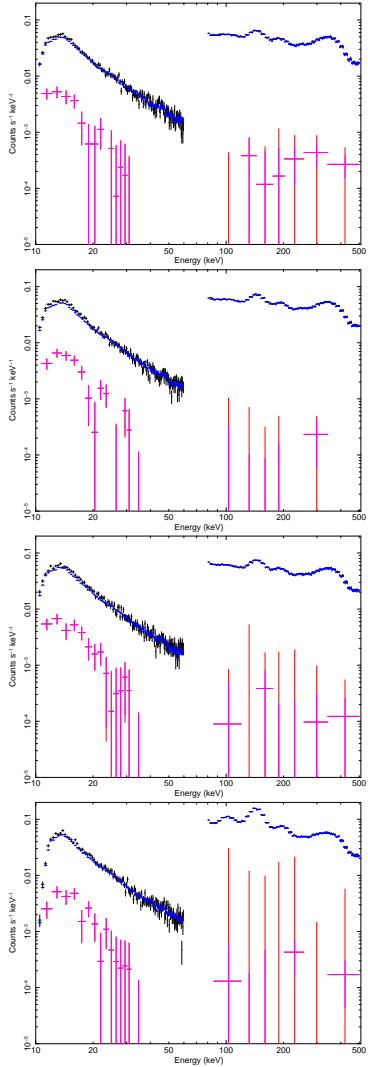

\centerline{\includegraphics[angle=-90,width=0.31\textwidth]{figure3a.ps}}
\centerline{\includegraphics[angle=-90,width=0.31\textwidth]{figure3b.ps}}
\centerline{\includegraphics[angle=-90,width=0.31\textwidth]{figure3c.ps}}
\centerline{\includegraphics[angle=-90,width=0.31\textwidth]{figure3d.ps}}
\caption{Same spectra as Figure \ref{fig:raw_spec_sn2014j}, 
but before OBS2014 (see Table \ref{table:observation_list}).
Panels from the top to bottom represent the X-ray spectra in
OBSID = 100033010, 100033020, 100033030, and 702026010, respectively. }
\label{fig:raw_spec_m82}
\end{figure}

The NXB-subtracted X-ray spectra in OBS2014 are shown 
in Figure \ref{fig:raw_spec_sn2014j}.
The systematic errors of NXB models for HXD PIN and GSO 
are included in the plots;
systematics for PIN NXB is set to be 3\% \citep{Fukazawa09} and
those for GSO is set to values in Table \ref{table:nxb_systematic_sn2014j}
determined by the short earth occultation data of this observation
as the worst cases.
Therefore, the GSO data in OBS2014 are still significant
in energy bins ({\it iv}), ({\it v}), ({\it vi}) 
in Table \ref{table:nxb_systematic_sn2014j},
whereas those in previous observations towards M82 are not significant 
as plotted in Figure \ref{fig:raw_spec_m82}.
In summary, we detected marginal signals from OBS2014 
in the 90 -- 500 keV band with about 
90\% confidence level (i.e., about $2 \sigma$).

\subsection{ULX and CXB contaminations}
\label{section:analysis:background}
We now discuss in more detail the GSO signals 
in the three energy bins ({\it iv}), ({\it v}), ({\it vi}) 
in Table \ref{table:nxb_systematic_sn2014j}, 
which corresponds to the 170 -- 350 keV band,
which turn out to be most significant in Figure \ref{fig:raw_spec_sn2014j}.
In these GSO energy bands, any possible SN2014J signal could be 
contaminated from 
the ULX M82 X-1 signal and Cosmic X-ray background (CXB) emission.

\begin{figure}[h]
\centerline{\includegraphics[angle=0,width=0.4\textwidth]{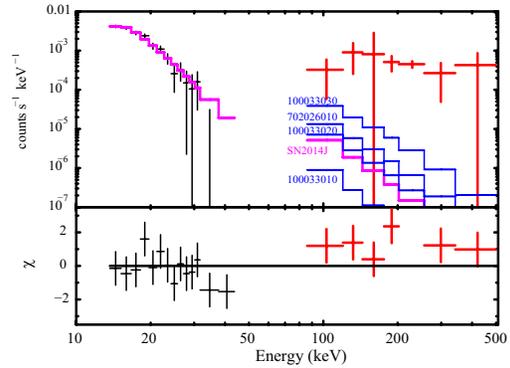}}
\caption{Top panel represents the same plots as 
red crosses of Figure \ref{fig:raw_spec_sn2014j}
(i.e., PIN and GSO spectra taken in OBS2014,
which are shown in black and red, respectively, in this plot), 
with the best-fit power-law model in magenta to reproduce the PIN data.
The bottom panel shows the chi values to this model.
Similarly, best-fit power-law models determined in other observations 
\citep[OBSID=100033010, 100033020, 100033030, and 702026010;][]{Miyawaki09}
are also plotted for GSO in blue.}
\label{fig:spec_sn2014j_ulx}
\end{figure}

The hard X-ray emission from M82 X-1 
can be estimated by the direct and simultaneous measurements 
with HXD PIN in the 13 -- 70 keV band.
The PIN spectrum in OBS2014 is well described by the single power law model,
which is usually used for a ULX \citep{Miyawaki09}.
The best fit model has a photon index of $3.93_{-0.40}^{+0.43}$ and 
an X-ray flux of $1.59_{-0.08}^{+0.06} \times 10^{-11}$ erg cm$^{-2}$ s$^{-1}$
in the 13 to 70 keV band with a reduced $\chi^2$ of 0.80 
under 12 degrees of freedom. 
Instead, the multi-color disk model \citep{Mitsuda84} is also 
used to represent the ULX spectra in several phases,
and is always below the power-law model in the harder X-ray band.
We therefore consider above power-law estimation as conservative,
and the value above corresponds to the upper limit 
of the contribution of M82 X-1 in the GSO band 
by an extrapolation from the best-fit power-law model in the PIN band.
In addition, the ULXs is usually variable \citep{Miyawaki09}
as is also seen in Figure \ref{fig:spec_sn2014j_ulx},
but the uncertainty on the flux from the PIN measurement here
is about 5 \%.
Therefore, the contamination from ULX is about $< 1.0 \pm 0.2 $ \% 
of the GSO signal in the 170--350 keV band.

\begin{figure}[h]
\centerline{\includegraphics[angle=0,width=0.35\textwidth]{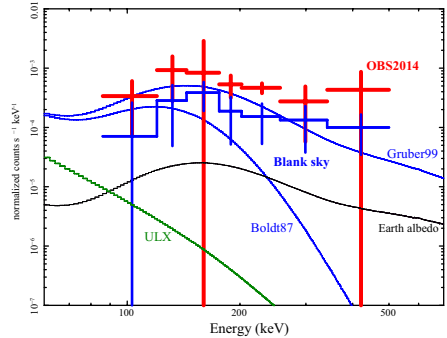}}
\caption{
Red crosses represent the NXB-subtracted GSO spectrum for OBS2014 
considering the Earth albedo emission in the NXB model (see the text).
The error bar includes statistical errors and systematic errors
of the NXB model and Earth albedo estimation.
The Cosmic X-ray backgrounds with HEAO-1 reported by 
\citet{Boldt87} and \citet{Gruber99} are shown in blue lines,
and the Earth albedo estimated for {\it Suzaku}
from the {\it Swift} and balloon experiments \citep{Ajello08,Imhof76} 
is shown in black line.
X-ray spectrum of GSO taken in blank sky observations
(listed in Table \ref{table:list_blanksky} with total exposure of 4.49 M sec)
is also plotted in blue crosses,
where the error bars contain systematic errors in the NXB model
and the Earth albedo.
Green line represents the ULX spectrum estimated by HXD PIN
(same as magenta line in Figure \ref{fig:spec_sn2014j_ulx}).
In conversion from CXB model into the data space, 
we used the GSO response matrix for a flat field emission,
accumulated effective areas in the Auxiliary-Response-File (ARF) database 
in CALDB (ae\_hxd\_gsoart\_20051126.fits).}
\label{fig:spec_sn2014j_cxb}
\end{figure}

In Figure \ref{fig:spec_sn2014j_cxb}, the GSO data of OBS2014 is 
compared with the canonical CXB model by HEAO-1 \citep{Gruber99},
which is confirmed with the recent hard X-ray observation with
{\it Swift} BAT \citep{Ajello08}.
For reference, another CXB model by \citet{Boldt87} is also plotted
but is not valid above the 100 keV band.
The CXB models were folded into the data space using the corresponding
detector's angular response, which is consistent with the estimation by
a full Monte Carlo simulation on {\it Suzaku} spacecraft 
with the Geant4 toolkit \citep{Terada05}.
The uncertainty on the angular response of GSO is checked 
by multiple pointing observation of Crab nebula \citep{Kokubun07}
but is not well derived yet. Therefore, we employ two alternatives; 
{\it a)} pulse height spectrum estimated from CXB model by \citet{Gruber99}
with the angular response, and
{\it b)} the hard X-ray spectrum with the HXD GSO
on the blank sky observations described in Section \ref {section:analysis:nxb}
In case {\it a)}, we put 10 \% uncertainty on the CXB spectral model
as described in \citet{Ajello08}.
As plotted in Figure \ref{fig:spec_sn2014j_cxb}, 
the X-ray flux in 200 -- 500 keV band of these two are consistent
with each other within $0.6 \sigma$ error,
whereas the latter tend to have harder spectral shape
(see Section \ref{section:discussion:cxb} for detail).
In the next section, we use both spectra for the CXB emission 
and then combine the two results to include systematic errors 
for the CXB estimation.

An additional systematic uncertainty may arise from
the contribution of the Earth albedo emission in the NXB model 
estimated from the Earth occultation data.
This is not considered in the current NXB model by \citet{Fukazawa09}.
The X-ray spectrum of the Earth albedo emission 
can be separated from the CXB spectra 
by changing the coverage of the Earth within the field of view
as has been done by the {\it Swift} BAT detector \citep{Ajello08}, 
but this method does not work for the HXD GSO in principle because of 
the design concept of the narrow field-of-view detector \citep{Takahashi07}.
Using the dependence of the Earth albedo level on the geomagnetic latitudes 
and the inclination angle $i$ of the spacecraft orbit to the Earth equator, 
the albedo for {\it Suzaku} at $i$ = 31 deg is simply interpolated 
between the {\it Swift} measurement \citep{Ajello08} at $i=20$ deg 
and balloon experiments at the polar and at the equator \citep{Imhof76}.
In this interpolation, we assume a systematic error of 25\%.
Such albedo emission in the NXB model contributes 
to increase a signal level compared with the CXB emission,
but at only about 10\% of the CXB level by \citet{Gruber99}, 
as plotted in Figure \ref{fig:spec_sn2014j_cxb}.
Therefore, this causes about 1 \% uncertainty for the GSO signal.

In summary, we have to subtract contributions of ULX and CXB emission
from the GSO signals and add the Earth's albedo to them 
of OBS2014 and the blank-sky observation (not to CXB models).
Numerically, the contributions of ULX, {\it a)} CXB \citep{Gruber99} or
{\it b)} blank sky spectrum, and the Earth albedo emission
to the NXB-subtracted GSO signals (albedo emission added) 
are 1\%, 49\%, 39\%, and 3\%, respectively, in the 170 -- 250 keV band.
Therefore, the GSO signal towards M82 in 2014 still remains 
at 4.0 or 2.5 $\sigma$ significance 
for case {\it a)} and {\it b)}, respectively,
in the 170 -- 250 keV band i.e., energy bins ({\it iv}) and ({\it v})
in Table \ref{table:nxb_systematic_sn2014j},
even after subtraction of the ULX and CXB emissions.

\subsection{Hard X-ray flux from SN2014J}
\label{section:analysis:flux}
\begin{figure}[h]
\centerline{\includegraphics[angle=0,width=0.34\textwidth]{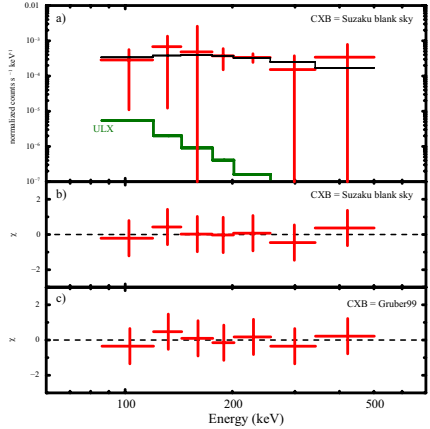}}
\caption{The top panel a) represents the X-ray spectrum of OBS2014
whose NXB and blank sky spectrum 
(blue crosses in Figure \ref{fig:spec_sn2014j_cxb}) are subtracted
and the Earth albedo emission (black line in Figure \ref{fig:spec_sn2014j_cxb})
is added. 
The error bars include the statistical errors and systematic errors
of the NXB model and Earth albedo estimation.
Contamination of ULX signals estimated by HXD PIN is shown in green
(same as green line in Figure \ref{fig:spec_sn2014j_cxb}),
which is fixed as a spectral model in the fitting.
Black line shows the best-fit power-law spectrum for the GSO data,
and the the chi values are plotted on the second panel b).
The bottom panel c) shows the same plot as b) but 
the HEAO-1 result by by \citet{Gruber99} is used 
for the subtraction of Cosmic X-ray background 
from the GSO data in the fitting, instead of the blank sky data.
In this fitting, the statistical errors and systematics
of NXB model, CXB model, and the Earth albedo estimations 
are considered.}
\label{fig:spec_sn2014j_fit}
\end{figure}

\begin{figure}[h]
\centerline{\includegraphics[angle=0,width=0.34\textwidth]{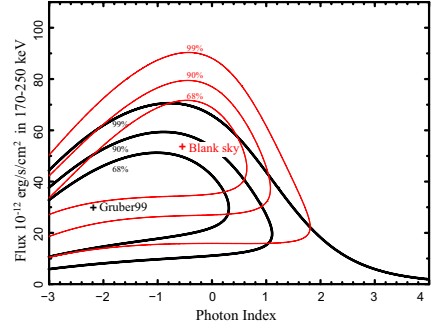}}
\caption{Confidence contour between the photon index and 
the X-ray flux in the 170 -- 250 keV band for the fitting of GSO data 
of OBS2014 in Figure \ref{fig:spec_sn2014j_fit}.
Red or black lines represent the contour from two fitting cases 
when the blank-sky data or HEAO-1 Model by \citet{Gruber99} 
are used as the Cosmic X-ray background level, respectively.
The + marks show the best fit values, and the the contours 
indicate the 68\%, 90\%, and 99\% levels from the inner to the outside.}
\label{fig:spec_sn2014j_fit_conf}
\end{figure}

In order to derive the X-ray flux from GSO signals numerically,
we performed spectral fittings with a power-law model on the GSO spectrum
after the subtraction of the NXB (Section \ref{section:analysis:nxb}) 
and the CXB with consideration of the Earth albedo 
(Section \ref{section:analysis:background}).
We tried two cases of CXB models 
(cases {\it a)} and {\it b)} in Section \ref{section:analysis:background})
to represent uncertainties of the CXB in the fitting.
The best fit models are shown in Figure \ref{fig:spec_sn2014j_fit}
and the hard X-ray flux in the 170 -- 250 keV band is found as 
$(0.9_{-0.3}^{+0.4})\times 10^{-4}$ ph s$^{-1}$ cm$^{-2}$ or
$(1.6 \pm 0.4)\times 10^{-4}$ ph s$^{-1}$ cm$^{-2}$ 
for cases {\it a)} and {\it b)} with $1 \sigma$ errors, respectively.
As shown in Figure \ref{fig:spec_sn2014j_fit_conf},
the normalization of the power-law model becomes zero
at 99\% significance level for case {\it a)}
and the significance of the measured signal is about 
90\% confidence level (i.e., $2 \sigma$) in total, 
as already described in section \ref{section:analysis:nxb}.
Therefore, we conclude that the {\it Suzaku} constain
the X-ray flux of SN2014J to below $2.2 \times 10^{-4}$ ph s$^{-1}$ cm$^{-2}$
at the 170 -- 250 keV band ($3 \sigma$ limit).

\section{Discussion}
\label{section:discussion}

\subsection{Detection of $\gamma-$rays with {\it Suzaku}}
\label{section:discussion:gamma_rays}

From the hard X-ray observation of SN2014J with {\it Suzaku} HXD
at $t = 77 \pm 2$ days after the explosion 
(Section \ref{section:observation}), 
the hard X-ray flux in the 170 -- 250 keV band is 
constrained with the 99\% ($3 \sigma$) upper limit of
$< 2.2 \times 10^{-4}$ ph s$^{-1}$ cm$^{-2}$.
This measurement complements the {\it INTEGRAL} measurements 
of soft X-ray band flux, at similar sensitivity obtained
with a shorter exposure.

\begin{figure}[h]
\centerline{\includegraphics[angle=0,width=0.4\textwidth]{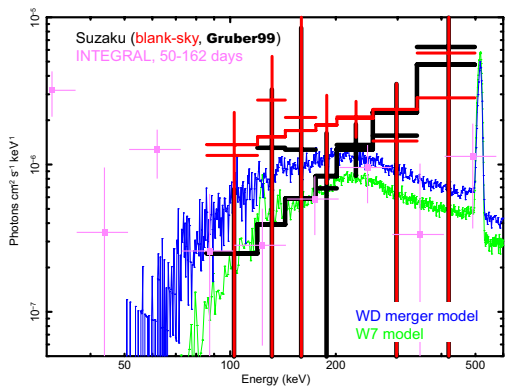}}
\caption{Red and black plots represent the photon spectra 
converted from the raw spectra in Figure \ref{fig:spec_sn2014j_fit}, 
assuming the power-law model with the best-fit values 
for the data whose CXB component are set to the blank sky data (see the text) 
or the spectral model by \citet{Gruber99}, respectively. 
Crosses and lines for them represent the data and the best-fit models, 
respectively.
For reference, photon spectra with the {\it INTEGRAL} ISGRI at 50 -- 162 days 
\citep[Figure 8 blue in ][]{Churazov15} is also plotted in magenta crosses.
Green and blue lines shows the spectral models of W7 \citep{Maeda12} and 
the white dwarf merger \citep{Summa13}, respectively, 
at the 75 days after the explosion with the $^{56}$Ni mass of $\sim 0.6 M_\odot$.}
\label{fig:photon_spectrum_and_models}
\end{figure}

The {\it Suzaku} upper limit at $77 \pm 2$ days is
consistent with those reported by {\it INTEGRAL} 
for the continuum emission in the 200 -- 400 keV band 
at $(2.0\pm0.8)\times 10^{-4}$ ph s$^{-1}$ cm$^{-2}$
at $75 \pm 25$ days
\footnote{the value by {\it INTEGRAL} is found 
only in the archive (astroph/1405.3332) of \citet{Churazov14}} within errors 
if we correct the energy width assuming a flat spectrum as indicated by
the spectral models of \citet{Maeda12}.
If we take the 68\% confidence levels 
(i.e., equivalent to the $1 \sigma$ errors)
in the systematic and statistical uncertainties of 
the {\it Suzaku} measurement,
the X-ray flux becomes $(1.2 \pm 0.7) \times 10^{-4}$ ph s$^{-1}$ cm$^{-2}$ 
in the same energy range.
This is consistent with {\it INTEGRAL} results within uncertainties.
The consistency can be found in Figure \ref {fig:photon_spectrum_and_models},
which shows the photon spectra estimated by the best-fit power-law models
in cases {\it a)} and {\it b)}, 
compared with the spectra by \citet{Churazov15}.

\subsection{Type Ia SN models}
\label{section:discussion:gamma_rays}

At $\sim 70$ days after the SN Ia explosion, 
decays of $^{56}$Co to $^{56}$Fe provide a major input 
into high energy radiation and thermal energy of the SN ejecta. 
The strongest lines are those at 847 keV and 1238 keV. 
The annihilation of positrons from this $\beta+$ decay
also produces either strong lines at or continuum
below $511$ keV. This high-energy radiation is degraded to lower
energy by Compton scattering, and below $\sim 200$ keV the photons are
absorbed by photoelectric absorption. These processes create characteristic
continuum emission from SNe Ia in the hard X-ray and soft gamma-ray regimes.

Figure \ref {fig:photon_spectrum_and_models} shows 
the photon spectrum obtained by the {\it Suzaku} observation.
This photon spectrum is constructed assuming a power law, 
and with the assumption of the best-fit power-law models 
either by {\it a)} the CXB model by \citet{Gruber99} or 
by {\it b)} the blank sky observations. 
In the same figure, the synthetic spectra of the W7 model \citep{Maeda12} 
and the violent merger model of a $0.9 M_\odot$ and a $1.1 M_\odot$ WD 
\citep{Summa13} are compared. 
In these models, the $^{56}$Ni-rich region, 
as well as the layers of intermediate-mass elements 
above the $^{56}$Ni-rich region, serve as the Compton-scattering layers. 
The W7 and delayed detonation models are (more or less) spherical, 
while the merger model has a large asymmetry 
in the distribution of the ejected material. 
In Figure \ref {fig:photon_spectrum_and_models}, 
we only show the angle-averaged model spectra; 
the viewing angle effect is considered later.
Both models have $M$($^{56}$Ni) $\sim 0.6 M_\odot$, 
which is consistent with what is inferred from optical properties 
(e.g., peak luminosity) of SN 2014J \citep{Ashall14}. 

The photon flux at 170-250 keV, taking our $2 \sigma$ signal, 
is indeed consistent with these models, 
within a systematic error related to the CXB. Above $\sim 300$ keV, 
the nominal flux level in the {\it Suzaku} spectrum is above 
the level of the CXB (for both CXB models),
leaving no residual SN2014J signal contribution, within uncertainties.

The most important difference in these two models is 
the total mass of the exploding system. 
The W7 model \citep{Nomoto82} is a representative of an explosion 
of a single Chandrasekhar-mass WD and 
the expected $\gamma$-ray emission is similar to 
other model variants such as deflagration-detonation models 
in a Chandrasekhar-mass \citep{Maeda12}. 
On the other hand, in the violent merger model both of the 
(sub-Chandrasekhar-mass) WDs are disrupted, 
leading to the super-Chandrasekhar mass for this particular model 
presented here \citep{Pakmor12,Roepke12,Summa13}. 
In terms of the expected $\gamma$-ray signals, 
the two models are characterized by different optical depth 
to $\gamma$-rays through Compton scattering. 
The violent merger model has more massive ejecta and thus is more opaque (by a factor of about two),
leading to a higher level of Compton continuum in the energy range 
of {\it Suzaku} observations. 
This difference is seen in Figure \ref {fig:photon_spectrum_and_models}. 

\begin{figure*}[t]
\centerline{
\includegraphics[width=0.8\textwidth]{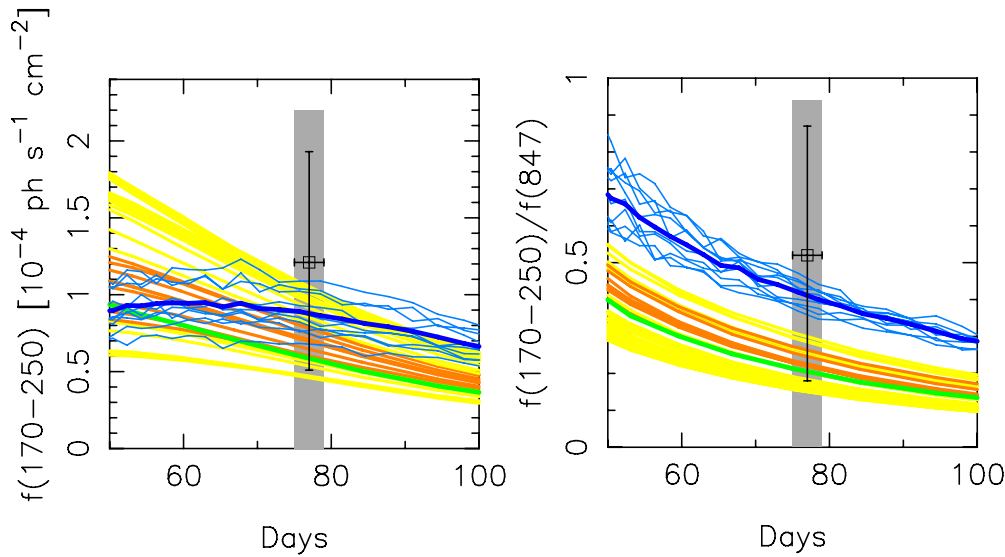}
}
\caption{The model light curves from some SN Ia models are compared 
with the {\it Suzaku} observations result. 
The upper panel shows the flux integrated in the energy of $170-250$ keV, 
while the lower panel shows the ratio of the $170-250$ keV flux to 
the 847 keV line flux. 
The $3 \sigma$ upper limit is shown by a gray area, while the flux of the marginal detection is shown by an open square with an error of $1 \sigma$. 
The models shown here are the W7 model \citep[green:][]{Nomoto82}, 
two-dimensional delayed-detonation models 
with various ignition conditions \citep[yellow and orange:][]{Maeda10,Maeda12},
and a violent merger of a $1.1 M_\odot$ and a $0.9 M_\odot$ WD 
\citep[cyan and blue:][]{Roepke12,Summa13}. 
For the violent merger model, 
the same model viewed from different directions (cyan) are shown 
together with an angle-averaged emission (blue). 
For the delayed-detonation models, 
the angle-variation is not large at these epochs, 
and only angle-averaged behaviors are shown for 32 models 
covering a range of $M$($^{56}$Ni). 
Models with $M$($^{56}$Ni) $= 0.47 - 072 M_\odot$, 
which are compatible to observational features of SN 2014J \citep{Ashall14}, 
are shown by orange curves, 
while the other models with larger/smaller $M$($^{56}$Ni) 
are shown by yellow curves.}
\label{fig:discuss_fluxratio}
\end{figure*}

Figure \ref{fig:discuss_fluxratio} shows the light curves 
integrated in the energy range of $170-250$ keV for various models, 
as well as the evolution of the ratio of the same continuum flux 
to the 847 keV line flux. 
The {\it Suzaku} upper limit and $1 \sigma$ data 
are also plotted as an one snapshot point at $t=77\pm2$ days 
thanks to the low-background capability of the HXD.
For the ratio, we took the flux from the {\it INTEGRAL} observation 
at $t=75\pm25$ days after the explosion \citep{Churazov14}. 
We adopt the energy range of $170-250$ keV since this corresponds to 
the marginally detected signal by {\it Suzaku} at $2 \sigma$. 
Shown in the figure are the W7 model, 
2D delayed detonation models \citep{Maeda12} and 
the violent merger model of \citet{Pakmor12}
for which gamma-ray observables have been presented
by \citet{Summa13}. 
The delayed detonation models were computed for different initial conditions 
(always with the assumption of a Chandrasekhar-mass WD),
covering a wide range of $M$($^{56}$Ni).
The models with $M$($^{56}$Ni) $= 0.47 - 0.72 M_\odot$ are indicated by 
yellow curves, and the W7 model and the violent merger model both have 
$\sim 0.6 M_{\odot}$ of $^{56}$Ni synthesized in the explosion. 
These are models compatible to the optical features of SN 2014J. 
The emission from the violent merger model is sensitive to 
the viewing angle even at $\sim 60 - 80$ days, 
and thus this is shown for various viewing angles.

It is seen that for a similar amount of $^{56}$Ni, 
the violent merger model having super-Chandrasekhar mass 
in the total ejecta, predicts a larger flux than the models 
with Chandrasekhar-mass WD progenitors. 
This is a result of the larger optical depth as explained above. 
Within the observational error, 
all the models are consistent with the {\it Suzaku} data. 

The difference between the violent merger model and 
the other explosion models becomes clearer
in the evolution of the flux ratio. 
Indeed, the ratio of the continuum flux to the line flux has been suggested 
to be a diagnostic to distinguish the progenitor WD mass \citep{Sim08,Summa13}  
-- In case of two models producing the similar amount of $^{56}$Ni, 
a larger amount of material surrounding the radioactive isotopes 
(for the larger WD mass) will convert a larger fraction of the line flux 
to the Compton down-scattered continuum flux. 
Therefore the ratio of the continuum to the line flux 
directly mirrors the ejecta mass.
This shows that it is in principle possible 
to constrain the total mass of the ejecta, 
thus the mass of the progenitor WD, through $\gamma$-ray observations. 
Unfortunately, the uncertainty in the {\it Suzaku} observation turned out 
to be too large to discriminate the models, 
even if we adopt the $1 \sigma$ error rather 
than the $3 \sigma$ upper limit. 
Unfortunately no constraint is obtained at a $3 \sigma$ level,
but it could already start to constrain some extreme models
while at a $1 \sigma$ level; 
the ratio predicted for some 2D delayed detonation models is 
below the {\it Suzaku} point beyond $1 \sigma$ error 
(i.e., the yellow lines below the data point in the lower panel of Figure 9), 
irrespective of the CXB model. 
All of these models have $M$($^{56}$Ni) $> 1 M_{\odot}$. 
These models have an extended distribution in $^{56}$Ni, 
and thus have small optical depths, leading to a low ratio. 
We thus reject, while at a $1 \sigma$ level, 
the models with such a large amount of $^{56}$Ni 
from the $\gamma$-ray signal alone 
fully independently of the optical emission. 

\subsection{CXB measurement with the HXD}
\label{section:discussion:cxb}
Very few models are reported for CXB emission in an energy band above 100 keV.
In Section \ref {section:analysis:background}, 
the hard X-ray spectrum with the HXD GSO in the 100 -- 500 keV band 
is presented in Figure \ref {fig:spec_sn2014j_cxb} 
and compared with the canonical CXB model by \citet{Gruber99}.
In the following fittings, the Earth's albedo emission 
estimated in Section \ref{section:analysis:nxb} is 
added to the NXB-subtracted spectrum of the blank sky observations.
Overall uncertainties on CXB model (10 \% from \citet{Ajello08}), 
angular response matrix (4 \% due to 
shade structure opaque to the sun in the X-ray mirror 
in Figure 11 of \citet{Terada05}), 
NXB estimation (0.19\% from Section \ref{section:analysis:nxb}),
and Earth's albedo emission (25\% from Section \ref{section:analysis:nxb}), 
are also included.
If we assume that the spectral shape of CXB is given by \citet{Gruber99},
the X-ray flux of the HXD/GSO blank-sky observation becomes 
$0.7 \pm 0.2$ times larger than the value of Gruber model.
Numerically, it is $(2.8\pm0.8) \times 10^{-2}$ ph s$^{-1}$ cm$^{-2}$ str$^{-1}$ 
or $(1.3\pm0.3) \times 10^{-8}$ erg s$^{-1}$ cm$^{-2}$ str$^{-1}$ 
in the 200 -- 500 keV band, where the errors represent statistics only.
If we reproduce the blank-sky spectrum with a simple power-law model, 
the photon index becomes harder than that of \citet{Gruber99} 
at $1.2^{+1.3}_{-1.0}$ and the X-ray flux becomes consistent with \citet{Gruber99} 
at $(5.1^{+2.5}_{-2.6}) \times 10^{-2}$ ph s$^{-1}$ cm$^{-2}$ str$^{-1}$ 
or $(2.6\pm1.3) \times 10^{-8}$ erg s$^{-1}$ cm$^{-2}$ str$^{-1}$ 
in the 200 -- 500 keV band, 
whereas the Gruber model corresponds to 
$4.1 \times 10^{-2}$ ph s$^{-1}$ cm$^{-2}$ str$^{-1}$ or 
$1.9 \times 10^{-8}$ erg s$^{-1}$ cm$^{-2}$ str$^{-1}$ in the same energy band.
Therefore, the X-ray spectrum of the blank sky observation with GSO 
reproduces the CXB model by \citet{Gruber99} within statistical errors.

\subsection{Future Perspectives}
\label{section:discussion:future}

A next-generation X-ray satellite {\it Hitomi} 
\citep[named {\it ASTRO-H} before launch;][]{AHTakahashi14}
has been successfully launched on 17 Feb 2016 and 
higher sensitivities than those of the HXD PIN/GSO or 
SPI/ISGRI on {\it INTEGRAL} will be achieved soon.
The background level of the soft gamma-ray detector 
\citep[SGD;][]{SGDTajima10,AHWatanabe12,AHFukazawa14} onboard {\it Hitomi} 
will be reduced by one order of magnitude compared to the HXD
and therefore soft $\gamma$-ray spectra from a future close-by Type-Ia SNe 
can be precisely measured as demonstrated in \citet{Maeda12}.
Thus, we can distinguish the explosion models 
between single and double degenerate progenitors 
as indicated in Figure \ref{fig:discuss_fluxratio}.
In distinctions of explosion models 
on Figure \ref{fig:discuss_fluxratio}, 
{\it Suzaku} demonstrated the importance of the snapshot measurement
achieving high sensitivity in a shorter exposure ($\pm 2$ days) 
than {\it INTEGRAL} ($\pm 25$ days).
In addition, we demonstrated in this paper that for future observations 
the refinement of the CXB spectral model is of critical importance.

\acknowledgments
{\bf Acknowledgments}\\
The authors would like to thank all the members of the {\it Suzaku} team 
for their continuous contributions in the maintenance of onboard instruments, 
spacecraft operation, calibrations, software development, 
and user support both in Japan and the United States; 
especially, we would like to thank the {\it Suzaku} managers 
for deep understandings on the importance of this ToO observation 
of SN2014J with {\it Suzaku} at the late stage of mission life.
The authors would like to thank 
H.~Sano, K.~Mukai, M.~Sawada, T.~Hayashi, T.~Yuasa, H.~Uchida, 
H.~Akamatsu for giving us private datasets of {\it Suzaku} observation 
in the NXB and CXB studies in Sections 
\ref{section:analysis:nxb} and \ref{section:analysis:background}.
This work was supported in part
by Grants-in-Aid for Scientific Research (B) from 
the Ministry of Education, Culture, Sports, Science and Technology (MEXT)
(No.~23340055 and No.~15H00773, Y.~T),
a Grant-in-Aid for Young Scientists (A) from MEXT
(No.~ 15K05107, A.~B.),
and 
a Grant-in-Aid for Young Scientists (B) from MEXT
(No.~ 26800100, K.~M.).
The work by K.M.\ is partly supported by World Premier
International Research Center Initiative (WPI Initiative), MEXT, Japan,
A.S. received support from the European Research Council through grant ERC-AdG
No. 341157-COCO2CASA, and FKR gratefully acknowledges 
the support of the Klaus Tschira Foundation.


{\it Facilities:} \facility{\it Suzaku}, \facility{\it INTEGRAL}, \facility{\it ASTRO-H}, \facility{\it Hitomi} 




\end{document}